\documentclass[fleqn,10pt]{wlscirep}
\usepackage[utf8]{inputenc}
\usepackage[T1]{fontenc}
\title{Topological carbon allotropes: paradigm shift for materials innovation}

\author[1,*]{Shinichi Saito}
\author[2]{Isao Tomita}
\affil[1]{Sustainable Electronic Technologies Research Group, Electronics and Computer Science, Faculty of Engineering and Physical Sciences, University of Southampton, SO17 1BJ, UK. }
\affil[2]{Department of Electrical and Computer Engineering, National Institute of Technology, Gifu College, 2236-2 Kamimakuwa, Motosu, Gifu 501-0495, Japan.}

\affil[*]{s.saito@soton.ac.uk}

\begin{abstract}
Topology is a central concept of mathematics, which allows us to distinguish two isolated rings with linked ones. 
In material science, researchers discovered topologically different carbon allotropes in a form of a cage a tube, and a sheet, which have unique translational and rotational symmetries, described by a crystallographic group theory, and the atoms are arranged at specific rigid positions in 3-dimensional ($D$) space.
However, topological orders must be robust against deformations, so that we can make completely different families of topological materials.
Here we propose various topological structures such as knots and links using covalent $\sigma$ bonds of carbon atoms, while allowing various topologically equivalent arrangements using weak $\pi$ bonds.
We found stable structures of topological molecules ($0D$) including Trefoil knots, a M$\bf \ddot{o}$bius strip, a Whitehead-link, and Borromean rings. 
In particular, we think a Hopf-link is especially important as a fundamental building block to construct Carbon-Nano-Chains ($1D$) and Chainmail ($2D$), which have topological-long-range-orders, tolerant against changes in geometries and chiralities.
By extending this idea, we invented a new 3D carbon allotrope, Hopfene, which has periodic arrays of Hopf-links to knit horizontal Graphene sheets into vertical ones without connecting by $\sigma$ bonds. 
\end{abstract}
\begin{document}

\flushbottom
\maketitle
%
%
\thispagestyle{empty}

\section*{Introduction}
Research areas on topology are spanning from pure mathematics\cite{Hopf31,Sunada12,Thurston82,Perelman03}, physics\cite{Berezinskii71,Kosterlitz72,Tanda02,Kane05,Hsieh08}, to chemistry\cite{Bissell94,Fang09,Molen09,Sauvage17,Dabrowski-Tumanski17}.
Mathematicians could even classify the shape of our universe\cite{Thurston82} and proved the Poincar$\rm \acute{e}$ conjecture successfully\cite{Perelman03}.
In condensed-matter physics, vortex and anti-vortex pairs in superconductors exhibit unusual topological phase transitions due to their binding and dissociations across the transition temperature\cite{Berezinskii71,Kosterlitz72}.
More recently, a topological insulator was proposed theoretically\cite{Kane05} and discovered experimentally\cite{Hsieh08}, which showed insulating character in the bulk regardless of its conducting surface. 
In these materials, quantum states exhibit inherent topological nature\cite{Berezinskii71,Kosterlitz72,Kane05} upon the appropriate choice of materials\cite{Hsieh08}.
On the other hand, nontrivial topological structures themselves were also chemically synthesised using polymers\cite{MacGillivray94,Carlucci03,Proserpio10}, macromolecules\cite{Bissell94,Fang09,Molen09,Sauvage17}, and proteins\cite{Dabrowski-Tumanski17}.
These pioneering works are well-established in soft materials\cite{MacGillivray94,Carlucci03,Proserpio10,Bissell94,Fang09,Molen09,Sauvage17,Dabrowski-Tumanski17}, and the scope of our present paper is to expand the concept to apply for hard materials made by strong covalent bonds while keeping the flexibility.

\subsection*{Topological molecules ($0D$)}
To start with, we considered many topological molecules (Methods, Figs. 1 and 2, Supplementary Videos 1 and 2).
The most direct applications of topological materials would be Nano-Electro-Mechanical-Systems (NEMS)\cite{Van18}, towards developing a molecular motor and associated advanced chemical technologies\cite{Bissell94,Tanda02,Fang09,Molen09,Sauvage17,Dabrowski-Tumanski17}. 
Figure 1a shows an ultimate example of trapped 1 benzene molecule using a straight-chain alkane of 1D atomistic carbon atoms connected to a Graphene sheet (Supplementary Video 1). 
It is technologically accessible to make such a molecular motor by combining patterning technologies of a free-standing Graphene sheet by ion or electron beam irradiations and growth technologies such as Chemical-Vapour-Deposition\cite{Kim09}.
The combination is very important to construct a topologically complex structures, because we must close the bond (in this case, the benzene ring) without affecting the other bonds (straight-chain).
Obviously, larger molecules and wider chain will be more suitable to construct a practical NEMS device.

In order to realise directional rotation of a molecular motor, it is important to control chirality of a molecule.
The most simplest nontrivial knot is the Trefoil knot, which has a chiral counter part (Fig. 1c).
The left (right) Trefoil knot has always left (right) spiral circulation as climbing up the strand (Supplementary Video 2).  
In other words, handedness is memorised in the molecule, which will be useful for applications in molecular memories and polarisation rotators. 

It is not possible to memorise the handedness solely by twisting a 1D carbon ring to the left or the right, since it merely change the geometry without changing the topology  (Fig. 1b).
In order to memorise the handedness, we need to lock the status of the twisted ring, by inserting the other molecule into the ring and close the bond. 
Then, we can make left and right topological molecules of Whitehead-links, whose handedness will be protected against the deformations of molecules as far as the bonds are sustained.

The situation is slightly changed, if we employ a Graphene nano-ribbon\cite{Nakada96}. 
Even if it consists of just several benzene rings, the ring will define a $2D$ plane to confine the ribbon so that it is different from a pure 1D chain.
As a result, the Möbius strip\cite{Tanda02,Flouris19} can memorise the handedness whether the Graphene nano-ribbons were twisted to the left or the right before bonding without using another Graphene nano-ribbon (Fig. 1d).
Moreover, the number of twisted rotations is also a topologically protected valuable as a winding number, which is also robust against the deformation.
We also made even more complex structure such as Borromean rings (Fig. 1e). 
To our surprise, the converged structure is rather beautiful with a proper symmetry upon exchanging rings, which was energetically favourable rather than keeping random deformations. 

Besides knots, the simplest nontrivial link is the Hopf-link\cite{Hopf31,Fang09,Sauvage17,Dabrowski-Tumanski17} (Fig. 1h-1o and Extended Fig. 2 ).
We found a benzene ring can sustain its $\sigma$ bonds, while allowing another benzene ring to penetrate without making a proper $\sigma$ bond.
The benzene ring can move freely without changing the topology of the Hopf-link, which will be a huge advantage to make flexible devices.
However, the bond lengths of the benzene-based Hopf-link are significantly expanded, so that the high tensile strains of the order of 20\% are accumulated (Extended Fig. 2a ).
Therefore, it is not possible to introduce 2 benzene rings into 1 ring (Fig. 2b and Supplementary Video 2). 
This will certainly limit the use of the benzene-based Hopf-link towards the construction of the $1D$ chain.

However, if we use 2 connected benzene rings (Naphthalene, C$_{10}$H$_{8}$), we can overcome this problem, since the insertion of just 1 benzene ring into the Naphthalene is enough for extending the length of the $1D$ Carbon-Nano-Chain (Fig. 2c and Fig. 3). 
We also considered to open-up one of the benzene ring of Naphthalene (Fig. 1 b), which corresponds to use a larger ring (Cyclodecapentaene, C$_{10}$H$_{10}$) for the Hopf-link (Fig. 2d), whose bonds are significantly relaxed.
In the larger ring, we could introduce as many as 4 rings (Fig. 2e), so that we can also construct the $2D$ 4-in-1 Chainmail (Fig. 3c) in addition to a standard Chain (Fig. 3b).

\subsection*{Topological Chains ($1D$) and Chainmail ($2D$)}

The unique features of these Chains and Chainmail are their flexibility with the rigidity ensured by the $\sigma$ bonds (Fig. 3 and Supplementary Video 3). 
Our Chains can be aligned to be straight-lines or to be rolled-up without breaking the topological links (Fig. 4).
The lengths can be changed to be several times, while it will be difficult to change the lengths of Carbon-Nano-Tubes\cite{Iijima91} due to its strong rigidity.
The $\sigma$ bonds using SP$_2$ orbitals are considered to be even stronger than SP$_3$ bonds of diamond, so that our chains will be much stronger than flexible polyacetylene\cite{Shirakawa77}. 
It is interesting to consider a similarity of our proposed Chain to a macroscopic metallic chain and chainmail, against the Fullerene\cite{Kroto85} to a football.
We think that it would not be mere coincidence that we came up to the same structure that the human race is already using for a long time.
The idea behind the football shape would be its high spherical symmetry.
It is the compatibility of flexibility with its rigidity why people made a chain, by introducing the Hopf-links\cite{Hopf31} even without being aware of mathematical rigidity.
The argument would also be true in the molecular level (Fig. 4). 

We have simulated various configurations of Chains by changing the boundary conditions. 
For example, by introducing a twisted boundary conditions between one end of the chain to the other for the $1D$ Hopf-linked Chain, we could introduce a global kink as a form of a soliton, which is a topologically protected excitation (Fig. 4).
Upon rotating twice, we found the chain was broken due to the high tensile strains accumulated in the Chain  (Fig. 4e).

The chain can memorise the global winding number and the chirality (the left or the right rotation) upon twisting the Chain (Fig. 4).
This situation is completely different from a $1D$ atomic chain like a polyacetylene (Fig. 4f), due to its rotational symmetry. 
One might think that polyacetylene would be more tolerant against the rotational disturbance, but no restoration force will work to prevent the rotation. 
On the other hand, the Hopf-linked chain can accumulate the global strain and the restoration force will be generated to protect itself against the structural deformation.
This would be related to the reason why DNA with a double-stranded spiral structure was naturally selected due to its stronger tolerance in a competitive environment. 

The geometrical coordinates of straight and rolled Carbon-Nano-Chains are apparently different, while both structures are topologically equivalent.
Therefore, we can regard these chains as possessing {\it Topological-Long-Range-Order (TLRO)} even without having translational nor rotational symmetries.
We can construct whole new varieties of materials with TLRO, as {\it topological materials}, which will not be categorised as a conventional crystal nor a completely-randomised amorphous material. 
Examples of Chains and Chainmail are clear examples of topological materials.

\subsection*{Topological Crystals ($3D$)}

The concept of {\it Topological-Long-Range-Order (TLRO)} is also applicable to a $3D$ structure, and the idea of a {\it topological crystal}\cite{MacGillivray94,Carlucci03,Proserpio10} would be as exciting as a time crystal\cite{Shapere12}.
We have conceived to the idea of a new $3D$ carbon allotrope by considering nested structures with stacked Graphene sheets both vertically and horizontally, while the intersections of these sheets are made of Hopf-links (Figa. 5 and 6, Supplementary Figs. 1-5, and Supplementary Video 4). 
To our surprise, the optimised structures have almost perfect translational symmetries, so that these will be realised as a conventional single crystal.

We propose to call this new structure, as {\it Hopfene}, named after a mathematician, Heinz Hopf, because of the importance of Hopf-links\cite{Hopf31} for this crystal. 
Similar to Carbon-Nano-Tubes\cite{Iijima91}, which has a family of allotropes depending on how to roll-up a Graphene sheet, Hopfene also has a family depending on how to insert Graphene sheets (Methods, Supplementary Figs. 1-5, and Supplementary Video 4).
Graphene sheets were inserted perpendicular to the other stacks, and it was important to align the directions of the zig-zag edges in parallel to match the periodicity of the Graphene lattice (Supplementary Fig. 2).
Let's assume that we inserted the first sheet to the bottom available slot, labelled as slot 0  (Supplementary Fig. 2).
If we insert next sheets into the most adjacent available slot, slot 1, for both horizontal ($x$) and vertical ($y$) directions, the structure is called as (1,1) Hopfene  (Figs. 5a, 5d, and Supplementary Fig. 2).
In this case, the adjacent stacks are half-lattice constant shifted ($c/2$) along zig-zag edge ($z$) direction, so that the stack is called as AB-stack.
On the other hand, if we insert Graphene sheets into slot 2, while making the slot 1 empty, the structure is called as (2,2) Hopfene  (Figs. 5b, 5e, and Supplementary Fig. 2).
In this case, the nearest adjacent sheets are aligned so that the stack is called as AA-stack.
We considered more complicated insertions, such as the insertion into slot 1 for $x$-direction, while the insertion along $y$-direction is from slot 2, whose structure is (1,2) Hopfene (Figs. 5c, 5f, and Supplementary Fig. 3).
In general, there exists $(n,m)$ Hopfene, where $n$ and $m$ are integers to describe the insertion of Graphene sheets, and if $n$ and $m$ are odd (even), the sheets are AB (AA) stack.
If $n=m$, the Hopfene structure is tetragonal with the lattice constants $a=b\neq c$, while it is orthogonal ($a\neq b\neq c$) for $n\neq m$. 
Therefore, we can recognise rectangular holes if we see the structure from the bottom (Fig. 5).
Therese holes would allow carrier doping by intercalation, which may lead superconductivity similar to alkali-metal doped Fullerenes\cite{Hebard91}.
If $n$ and $m$ are large, the size of the hole will be large, so that the Graphene sheets and the crystal itself can be deformed substantially, breaking the translational symmetry. 
It is also expected that the actual experimental structures would be different due to the huge strains accumulated in rings, as theoretically predicted in a highly-strained Graphene, which stabilised to be a dimerised Kekul$\rm \acute{e}$-like structure\cite{Sorella18}.
Even in that case, TLRO would be kept as far as $\sigma$ bonds survived to keep Hopf-links.

It is straightforward to extend this concept\cite{MacGillivray94,Carlucci03,Proserpio10} to make {\it heterogeneous topological structures} by introducing different materials such as boron-nitride and molybdenum-disulfide. 
The unique aspect of this approach is a topological link to bind various sheets strongly together without forming a proper chemical bond.
This configuration is {\it topologically different} from the simple stacking weakly bound together by van der Waals force\cite{Koma84}. 

As the first step towards making Hopfene, we propose to make Hopf-linked bilayer-Graphene (Fig. 7 and Supplementary Fig. 6).
Unlike to the stacked bilayer-Graphene\cite{Oshima97,Novoselov04,Ando05,Ferrari06}, the Graphene sheets are linked only at the $1D$ chain, so that the impacts of the coupling on the band structure would be limited.
Reflecting the double layers, momentum space will also be double, leading to the crossed layers even in momentum space (Fig. 7b).
The valleys are degenerate at the same points, because the direction of the zig-zag edge ($z$) is the same for both layers due to sharing of the $1D$ chain. 
As we increase the number of inserted Graphene sheets, momentum space will be eventually filled by layers towards the complete $3D$ band structure.
It is beyond the scope of this paper to predict how the $2D$ Dirac Fermions\cite{Kane05,Ando05,Nakada96,Sorella18,Armitage18} are crossing over to the excitations in $3D$ Hopfene.

\section*{Conclusion}
We have validated the concept of topological materials\cite{MacGillivray94,Carlucci03,Proserpio10} in materials with hard covalent bonds, which we believe, will be discovered for the future. 
$1D$ Carbon-Nano-Chains, Hopf-linked bilayer-Graphene ($2D$), and $3D$ Hopfene have been proposed as examples of new {\it topological carbon allotropes}, which will be useful to examine fundamental physics of massless Dirac Fermions\cite{Kane05,Ando05,Nakada96,Sorella18,Armitage18} in topologically nontrivial geometries.
We envisage various practical applications (Extended Fig. 12) of these hard and flexible topological materials for DNA-sensing, functional materials, and NEMS.

\section*{Methods}

We have used Chem3D, a molecular editor, for all our simulations\cite{Evans14}.

For the simulations of the rolled Chains (Figs. 3 and 4) and Chainmail (Fig. 3), we have gradually changed the initial coordination of atoms.
If we changed the initial condition significantly different from the previously converged geometries, Hopf-links were completely broken and the structures were topologically changed.
This will also happen in the real situation, where the external force exceeded the threshold enough to break one of the $\sigma$ bonds for the Hopf-links.
Conversely, the proposed Chains and Chainmail will be topologically robust as far as the $\sigma$ bonds are maintained.
Thus, we expect a topologically flexible material, which would be as flexible as a rubber, while it is very stiff as hard as or even harder than a diamond due to its SP$_2$ nature of the rings. 

Hopfene was made by preparing AA-stacked Graphene sheets, separated by the distance of the lattice constant of a Graphene, which is the adjacent distance between the sides of the ring (Supplementary Figs. 4 and 5).
Depending on the targeted stacking, described by $(n,m)$ where $n$ and $m$ are integers for the available slots along horizontal ($x$) and vertical ($y$) directions, the number of inserted Graphene sheets and separations were adjusted.
We have prepared the Graphene sheets by hands using the GUI (Graphical User Interface) of the software, so that the position and the distance were not perfect.
This uncertainty resulted in the formation of {\it  topological defects} in the final structure (Supplementary Fig. 1), but we kept as it is to highlight new types of defects without broken bonds.
Then, we optimised the structure relying on the algorithm of the software to minimise the energy.
After the optimisation, for (1,1) Hopfene, AB-stack was automatically chosen for both $x$ and $y$ directions by shifting the Graphene sheets with the amount of the half of the lattice constant along $z$ direction, regardless of the initial condition of AA-stack.
Here, please note that AB-stack of Hopfene is different from AB-stack of Graphite.
AB-stack of Hopfene means the parallel-shift of the Graphene sheet along the zig-zag edge (Fig. 6), while AB-stack of Graphite is the parallel-shift along the arm-chair edge, so that the directions are perpendicular to that of Hopfene.
The process of the convergence of the structure would be similar to the growth process in experiments for the future, so that we have recorded as a video (Supplementary Video 4).


\begin{thebibliography}{10}
\urlstyle{rm}
\expandafter\ifx\csname url\endcsname\relax
  \def\url#1{\texttt{#1}}\fi
\expandafter\ifx\csname urlprefix\endcsname\relax\def\urlprefix{URL }\fi
\expandafter\ifx\csname doiprefix\endcsname\relax\def\doiprefix{DOI: }\fi
\providecommand{\bibinfo}[2]{#2}
\providecommand{\eprint}[2][]{\url{#2}}

\bibitem{Hopf31}
\bibinfo{author}{Hopf, H.}
\newblock \bibinfo{journal}{\bibinfo{title}{{\"U}ber die abbildungen der
  dreidimensionalen sph{\"a}re auf die kugelfl{\"a}che}}.
\newblock {\emph{\JournalTitle{Mathematische Annalen}}}
  \textbf{\bibinfo{volume}{104}}, \bibinfo{pages}{637--665},
  \doiprefix\url{10.1007/BF01457962} (\bibinfo{year}{1931}).

\bibitem{Sunada12}
\bibinfo{author}{Sunada, T.}
\newblock \bibinfo{journal}{\bibinfo{title}{Lecture on topological
  crystallography}}.
\newblock {\emph{\JournalTitle{Japan. J. Math.}}} \textbf{\bibinfo{volume}{7}},
  \bibinfo{pages}{1--39}, \doiprefix\url{10.1007/s11537-012-1144-4}
  (\bibinfo{year}{2012}).

\bibitem{Thurston82}
\bibinfo{author}{Thurston, W.~P.}
\newblock \bibinfo{journal}{\bibinfo{title}{Three dimentional manifolds,
  {K}leinian groups and hyperbolic geometry}}.
\newblock {\emph{\JournalTitle{Bull. Amer. Math. Soc. (N.S.)}}}
  \textbf{\bibinfo{volume}{6}}, \bibinfo{pages}{357--381},
  \doiprefix\url{projecteuclid.org/euclid.bams/1183548782}
  (\bibinfo{year}{1982}).

\bibitem{Perelman03}
\bibinfo{author}{Perelman, G.}
\newblock \bibinfo{journal}{\bibinfo{title}{Finite extinction time for the
  solutions to the ricci flow on certain three-manifolds}}.
\newblock {\emph{\JournalTitle{arXiv:math/0307245 [math.DG]}}}
  (\bibinfo{year}{2003}).

\bibitem{Berezinskii71}
\bibinfo{author}{Berezinskii, V.~L.}
\newblock \bibinfo{journal}{\bibinfo{title}{Destruction of long-range order in
  one-dimensional and two-dimensional systems having a continuous symmetry
  group i. classical systems}}.
\newblock {\emph{\JournalTitle{Soviet Phys. JETP}}}
  \textbf{\bibinfo{volume}{32}}, \bibinfo{pages}{493--500}
  (\bibinfo{year}{1971}).

\bibitem{Kosterlitz72}
\bibinfo{author}{Kosterlitz, J.~M.} \& \bibinfo{author}{Thouless, D.~J.}
\newblock \bibinfo{journal}{\bibinfo{title}{Long range order and metastability
  in two dimensional solids and superfluids}}.
\newblock {\emph{\JournalTitle{J. Phys. C: Solid State Phys.}}}
  \textbf{\bibinfo{volume}{5}}, \bibinfo{pages}{L124--L126}
  (\bibinfo{year}{1972}).

\bibitem{Tanda02}
\bibinfo{author}{Tanda, S.} \emph{et~al.}
\newblock \bibinfo{journal}{\bibinfo{title}{A m$\rm \ddot{o}$bius strip of
  single crystals}}.
\newblock {\emph{\JournalTitle{Nature}}} \textbf{\bibinfo{volume}{417}},
  \bibinfo{pages}{397--398}, \doiprefix\url{10.1038/417397a}
  (\bibinfo{year}{2002}).

\bibitem{Kane05}
\bibinfo{author}{Kane, C.~L.} \& \bibinfo{author}{Mele, E.~J.}
\newblock \bibinfo{journal}{\bibinfo{title}{$z_2$ topological order and the
  quantum spin hall effect}}.
\newblock {\emph{\JournalTitle{Phys. Rev. Lett.}}}
  \textbf{\bibinfo{volume}{95}}, \bibinfo{pages}{146802},
  \doiprefix\url{10.1103/PhysRevLett.95.146802} (\bibinfo{year}{2005}).

\bibitem{Hsieh08}
\bibinfo{author}{Hsieh, D.} \emph{et~al.}
\newblock \bibinfo{journal}{\bibinfo{title}{A topological {D}irac insulator in
  a quantum spin {H}all phase}}.
\newblock {\emph{\JournalTitle{Nature}}} \textbf{\bibinfo{volume}{452}},
  \bibinfo{pages}{970--974}, \doiprefix\url{10.1038/nature06843}
  (\bibinfo{year}{2008}).

\bibitem{Bissell94}
\bibinfo{author}{Bissell, R.~A.}, \bibinfo{author}{C{\'o}rdova, E.},
  \bibinfo{author}{Kaifer, A.~E.} \& \bibinfo{author}{Stoddart, J.~F.}
\newblock \bibinfo{journal}{\bibinfo{title}{A chemically and electrochemically
  switchable molecular shuttle}}.
\newblock {\emph{\JournalTitle{Nature}}} \textbf{\bibinfo{volume}{369}},
  \bibinfo{pages}{133--137}, \doiprefix\url{10.1038/369133a0}
  (\bibinfo{year}{1994}).

\bibitem{Fang09}
\bibinfo{author}{Fang, L.} \emph{et~al.}
\newblock \bibinfo{journal}{\bibinfo{title}{Mechanically bonded
  macromolecules}}.
\newblock {\emph{\JournalTitle{Chem. Soc. Rev.}}}
  \textbf{\bibinfo{volume}{39}}, \bibinfo{pages}{17--29},
  \doiprefix\url{10.1039/b917901a} (\bibinfo{year}{2010}).

\bibitem{Molen09}
\bibinfo{author}{v.~d. Molen, S.~J.} \emph{et~al.}
\newblock \bibinfo{journal}{\bibinfo{title}{Light-controlled conductance
  switching of ordered metal-molecule-metal devices}}.
\newblock {\emph{\JournalTitle{Nano Lett.}}} \textbf{\bibinfo{volume}{9}},
  \bibinfo{pages}{76--80}, \doiprefix\url{10.1021/nl802487j}
  (\bibinfo{year}{2009}).

\bibitem{Sauvage17}
\bibinfo{author}{Sauvage, J.~P.}
\newblock \bibinfo{journal}{\bibinfo{title}{From chemical topology to molecular
  machines ({N}obel lecture)}}.
\newblock {\emph{\JournalTitle{Angew. Chem. Int. Ed.}}}
  \textbf{\bibinfo{volume}{56}}, \bibinfo{pages}{11080--11093},
  \doiprefix\url{10.1002/anie.201702992} (\bibinfo{year}{2017}).

\bibitem{Dabrowski-Tumanski17}
\bibinfo{author}{Dabrowski-Tumanski, P.} \& \bibinfo{author}{Sulkowska, J.~I.}
\newblock \bibinfo{journal}{\bibinfo{title}{Topological knots and links in
  proteins}}.
\newblock {\emph{\JournalTitle{PNAS}}} \textbf{\bibinfo{volume}{114}},
  \bibinfo{pages}{3415--3420}, \doiprefix\url{10.1073/pnas.1615862114}
  (\bibinfo{year}{2017}).

\bibitem{MacGillivray94}
\bibinfo{author}{MacGillivray, L.~R.}, \bibinfo{author}{Subramanian, S.} \&
  \bibinfo{author}{Zaworotko, M.~J.}
\newblock \bibinfo{journal}{\bibinfo{title}{Interwoven two- and
  three-dimensional coordination polymers through self-assembly of {C}u'
  cations with linear bidentate ligands}}.
\newblock {\emph{\JournalTitle{J. Chem. Soc., Chem. Commun.}}}
  \textbf{\bibinfo{volume}{0}}, \bibinfo{pages}{1325--1326},
  \doiprefix\url{10.1039/C39940001325} (\bibinfo{year}{1994}).

\bibitem{Carlucci03}
\bibinfo{author}{Carlucci, L.}, \bibinfo{author}{Ciani, G.} \&
  \bibinfo{author}{Proserpio, D.~M.}
\newblock \bibinfo{journal}{\bibinfo{title}{Polycatenation, polythreading and
  polyknotting in coordination network chemistry}}.
\newblock {\emph{\JournalTitle{Coordin Chem Rev.}}} \bibinfo{pages}{247--289},
  \doiprefix\url{10.1016/S0010-8545(03)00126-7} (\bibinfo{year}{2003}).

\bibitem{Proserpio10}
\bibinfo{author}{Proserpio, D.~M.}
\newblock \bibinfo{journal}{\bibinfo{title}{Polycatenation weaves a 3d web}}.
\newblock {\emph{\JournalTitle{Nature Chem.}}} \textbf{\bibinfo{volume}{2}},
  \bibinfo{pages}{435--436}, \doiprefix\url{10.1038/nchem.674}
  (\bibinfo{year}{2010}).

\bibitem{Van18}
\bibinfo{author}{Van, N.~H.}, \bibinfo{author}{Muruganathan, M.},
  \bibinfo{author}{Kulothungan, J.} \& \bibinfo{author}{Mizuta, H.}
\newblock \bibinfo{journal}{\bibinfo{title}{Fabrication of a three-terminal
  graphene nanoelectromechanical switch using two-dimensional materials}}.
\newblock {\emph{\JournalTitle{Nanoscale}}} \textbf{\bibinfo{volume}{10}},
  \bibinfo{pages}{12349}, \doiprefix\url{10.1039/c7nr08439k}
  (\bibinfo{year}{2018}).

\bibitem{Kim09}
\bibinfo{author}{Kim, K.~S.} \emph{et~al.}
\newblock \bibinfo{journal}{\bibinfo{title}{Large-scale pattern growth of
  graphene films for stretchable transparent electrodes}}.
\newblock {\emph{\JournalTitle{Nature}}} \textbf{\bibinfo{volume}{457}},
  \bibinfo{pages}{706--710}, \doiprefix\url{10.1038/nature07719}
  (\bibinfo{year}{2009}).

\bibitem{Nakada96}
\bibinfo{author}{Nakada, K.}, \bibinfo{author}{Fujita, M.},
  \bibinfo{author}{Dresselhaus, G.} \& \bibinfo{author}{Dresselhaus, M.~S.}
\newblock \bibinfo{journal}{\bibinfo{title}{Edge state in graphene ribbons:
  Nanometer size effect and edge shape dependence}}.
\newblock {\emph{\JournalTitle{Phys. Rev. B}}} \textbf{\bibinfo{volume}{54}},
  \bibinfo{pages}{17954--17961} (\bibinfo{year}{1996}).

\bibitem{Flouris19}
\bibinfo{author}{Flouris, K.}, \bibinfo{author}{Jimenez, M.~M.} \&
  \bibinfo{author}{Herrmann, H.~J.}
\newblock \bibinfo{journal}{\bibinfo{title}{Quantum spin-{H}all effect on
  {M}{\"o}bius graphene ribbon}}.
\newblock {\emph{\JournalTitle{arXiv:1902.03892 [cond-mat.mes-hall]
  arXiv:1902.03892 [cond-mat.mes-hall]}}}  (\bibinfo{year}{2019}).

\bibitem{Iijima91}
\bibinfo{author}{Iijima, S.}
\newblock \bibinfo{journal}{\bibinfo{title}{Helical microtubules of graphitic
  carbon}}.
\newblock {\emph{\JournalTitle{Nature}}} \textbf{\bibinfo{volume}{354}},
  \bibinfo{pages}{56--58}, \doiprefix\url{10.1038/354056a0}
  (\bibinfo{year}{1991}).

\bibitem{Shirakawa77}
\bibinfo{author}{Shirakawa, H.}, \bibinfo{author}{Louis, E.~J.},
  \bibinfo{author}{Macdirmid, A.~G.}, \bibinfo{author}{Chiang, C.~K.} \&
  \bibinfo{author}{Heeger, A.~J.}
\newblock \bibinfo{journal}{\bibinfo{title}{Synthesis of electrically
  conducting organic polymers: Halogen derivatives of polyacetylene,
  ({CH})$_x$}}.
\newblock {\emph{\JournalTitle{J. C. S. Chem. Comm.}}}
  \bibinfo{pages}{578--580}, \doiprefix\url{10.1039/C39770000578}
  (\bibinfo{year}{1977}).

\bibitem{Kroto85}
\bibinfo{author}{Kroto, H.~W.}, \bibinfo{author}{Heath, J.~R.},
  \bibinfo{author}{O'Brien, S.~C.}, \bibinfo{author}{Curl, R.~F.} \&
  \bibinfo{author}{Smalley, R.~E.}
\newblock \bibinfo{journal}{\bibinfo{title}{C$_{60}$: Backminsterfullerene}}.
\newblock {\emph{\JournalTitle{Nature}}} \textbf{\bibinfo{volume}{318}},
  \bibinfo{pages}{162--163}, \doiprefix\url{10.1038/318162a0}
  (\bibinfo{year}{1985}).

\bibitem{Shapere12}
\bibinfo{author}{Shapere, A.} \& \bibinfo{author}{Wilczek, F.}
\newblock \bibinfo{journal}{\bibinfo{title}{Classical time crystals}}.
\newblock {\emph{\JournalTitle{Phys. Rev. Lett.}}}
  \textbf{\bibinfo{volume}{109}},
  \doiprefix\url{10.1103/PhysRevLett.109.160402} (\bibinfo{year}{2012}).

\bibitem{Hebard91}
\bibinfo{author}{Hebard, A.~F.} \emph{et~al.}
\newblock \bibinfo{journal}{\bibinfo{title}{Superconductivity at 18 {K} in
  potassium-doped {C}$_{60}$}}.
\newblock {\emph{\JournalTitle{Nature}}} \textbf{\bibinfo{volume}{350}},
  \bibinfo{pages}{600--601}, \doiprefix\url{10.1038/350600a0}
  (\bibinfo{year}{1991}).

\bibitem{Sorella18}
\bibinfo{author}{Sorella, S.} \emph{et~al.}
\newblock \bibinfo{journal}{\bibinfo{title}{Correlation-driven dimerization and
  topological gap opening in isotropically strained graphene}}.
\newblock {\emph{\JournalTitle{Phys. Rev. Lett.}}}
  \textbf{\bibinfo{volume}{121}}, \bibinfo{pages}{066402},
  \doiprefix\url{10.1103/PhysRevLett.121.066402} (\bibinfo{year}{2018}).

\bibitem{Koma84}
\bibinfo{author}{Koma, A.}, \bibinfo{author}{Sunouchi, K.} \&
  \bibinfo{author}{Miyajima, T.}
\newblock \bibinfo{journal}{\bibinfo{title}{Fabrication of ultrathin
  heterostructures with van der {W}aals epitacy}}.
\newblock {\emph{\JournalTitle{J Vac Sci Technol B}}}
  \textbf{\bibinfo{volume}{3}}, \doiprefix\url{10.1116/1.583125}
  (\bibinfo{year}{1985}).

\bibitem{Oshima97}
\bibinfo{author}{Oshima, C.} \& \bibinfo{author}{Nagashima, A.}
\newblock \bibinfo{journal}{\bibinfo{title}{Ultra-thin epitaxial films of
  graphite and hexagonal boron nitride on solid surfaces}}.
\newblock {\emph{\JournalTitle{J. Phys.: Condens. Metter}}}
  \textbf{\bibinfo{volume}{9}}, \bibinfo{pages}{1--20},
  \doiprefix\url{10.1088/0953-8984/9/1/004} (\bibinfo{year}{1997}).

\bibitem{Novoselov04}
\bibinfo{author}{Novoselov, K.~S.} \emph{et~al.}
\newblock \bibinfo{journal}{\bibinfo{title}{Electric field effect in atomically
  thin carbon films}}.
\newblock {\emph{\JournalTitle{Science}}} \textbf{\bibinfo{volume}{306}},
  \bibinfo{pages}{666--669}, \doiprefix\url{10.1126/science.1102896}
  (\bibinfo{year}{2004}).

\bibitem{Ando05}
\bibinfo{author}{Ando, T.}
\newblock \bibinfo{journal}{\bibinfo{title}{Theory of electronic states and
  transport in carbon nanotubes}}.
\newblock {\emph{\JournalTitle{J. Phys. Soc. Jpn.}}}
  \textbf{\bibinfo{volume}{74}}, \bibinfo{pages}{777--817},
  \doiprefix\url{10.1143/JPSJ.74.777} (\bibinfo{year}{2005}).

\bibitem{Ferrari06}
\bibinfo{author}{Ferrari, A.~C.} \emph{et~al.}
\newblock \bibinfo{journal}{\bibinfo{title}{Raman spectrum of graphene and
  graphene layers}}.
\newblock {\emph{\JournalTitle{Phys. Rev. Lett.}}}
  \textbf{\bibinfo{volume}{97}}, \bibinfo{pages}{187401},
  \doiprefix\url{10.1103/PhysRevLett.97.187401} (\bibinfo{year}{2006}).

\bibitem{Armitage18}
\bibinfo{author}{Armitage, N.~P.}, \bibinfo{author}{Mele, E.~J.} \&
  \bibinfo{author}{Vishwanath, A.}
\newblock \bibinfo{journal}{\bibinfo{title}{Weyl and {D}irac semimetals in
  three-dimensional solids}}.
\newblock {\emph{\JournalTitle{Rev. Mod. Phys.}}}
  \textbf{\bibinfo{volume}{90}}, \doiprefix\url{10.1103/RevModPhys.90.015001}
  (\bibinfo{year}{2018}).

\bibitem{Evans14}
\bibinfo{author}{Evans, D.~A.}
\newblock \bibinfo{journal}{\bibinfo{title}{History of the {H}arvard
  {C}hem{D}raw project}}.
\newblock {\emph{\JournalTitle{Angew. Chem. Int. Ed.}}}
  \textbf{\bibinfo{volume}{53}}, \bibinfo{pages}{11140--11145},
  \doiprefix\url{10.1002/anie.201405820} (\bibinfo{year}{2014}).

\end{thebibliography}

\section*{Acknowledgements}
This work is supported by EPSRC Manufacturing Fellowship (EP/M008975/1).
We would like to thank Prof. H. Mizuta, Dr. M. Muruganathan, Prof. Y. Oshima,  Prof. S. Matsui, Prof. S. Ogawa, Prof. S. Kurihara, and Prof. H. N. Rutt for stimulating discussions.
S.S also would like to thank JAIST for their hospitalities during his stay at the Center for Single Nanoscale Innovative Devices.

\section*{Author contributions statement}
S.S. and I.T. conceived the idea of topological carbon allotropes, together. 
S.S. conceived the simulations, and I.T. analysed the results. 
S.S. wrote the draft. 
S.S. and I.T.  reviewed the manuscript. 

\section*{Additional information}

\textbf{Accession codes} (The data from the paper can be obtained from the University of Southampton ePrint research repository: \\ 
http://dx.doi.org/10.5258/SOTON/D0884); 
\textbf{Competing interests} (The authors declare no competing interests both financially and non-financially.).

\pagebreak
\begin{figure}
\begin{center}
\includegraphics[width=12cm]{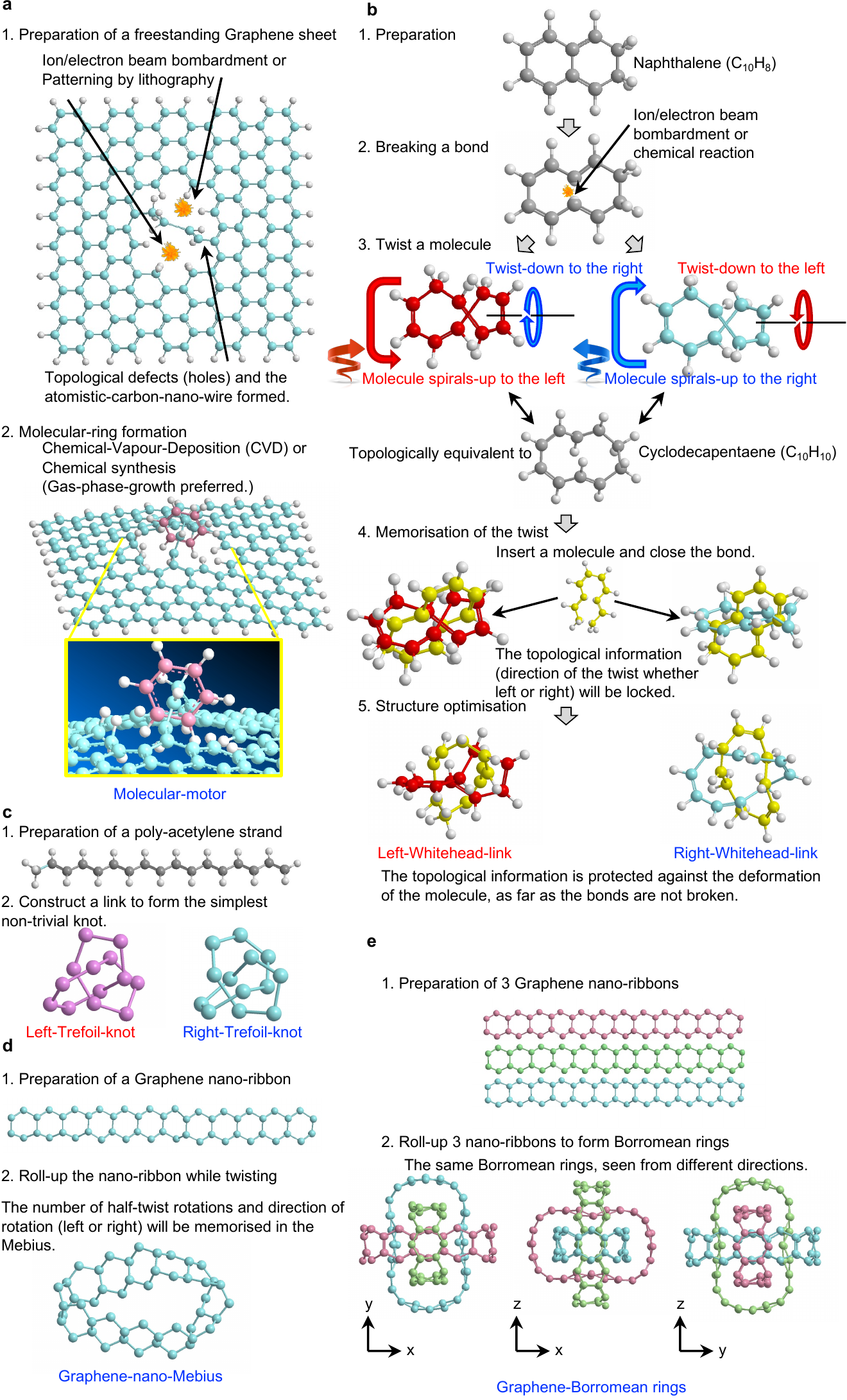}
\end{center}
\end{figure}
\pagebreak
\setcounter{figure}{0}
\begin{figure}
\begin{center}
\caption{
	{\bf  Proposed procedure to make topological structures.} 
	{\bf a}, Benzen-based molecular motor. A Graphene sheet is prepared, and patterned to form an atomic wire using ion/electron beam or lithography. Then, a benzene or other organic molecule is trapped in the patterned atomic wire.
	{\bf b}, Whitehead-links. 
Left and right links will be created by twisting molecules and locked by another atomic wires.
	{\bf c}, Trefoil-knots. Left and right knots can be created depending on how to connect the atomic wire.  
	{\bf d}, Möbius strip using a Graphene nano-ribbon. 
	{\bf e}, Borromean rings using Graphene nano-ribbons. 
}
\end{center}
\end{figure}

\pagebreak
\begin{figure}
\begin{center}
\includegraphics[width=18.3cm]{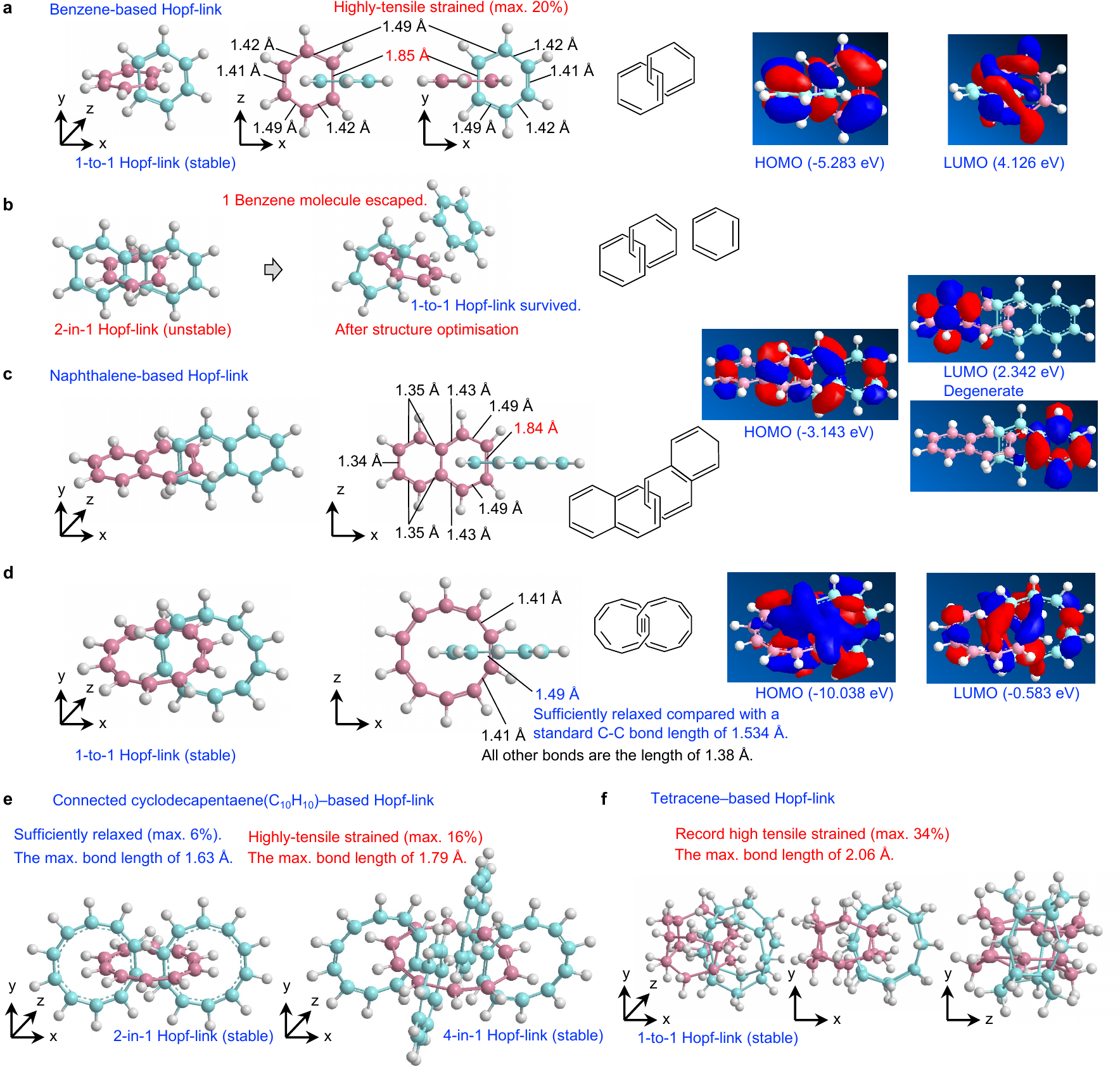}
\end{center}
\end{figure}
\pagebreak
\setcounter{figure}{1}
\begin{figure}
\begin{center}
\caption{
	{\bf  Hopf-linked molecules.} 
	{\bf a}, Benzen-based Hopf-linked molecule. 
The bonds penetrating into the other benzene ring were highly-strained.
The linked benzene rings are intersecting perpendicular to each other, and there are energetically-equivalent 6 positions corresponding to the sides of each ring, so that the total $6 \times 6=36$ configurations are allowed corresponding to its hexagonal structure.
Wavefunctions for both HOMO and LUMO are spreading to the entire link, suggesting the strong hybridisation of 2 rings.
	{\bf b}, Break-down of benzen-based Hopf-linked molecule.  A benzene ring can accommodate only 1 benzene-ring.
	{\bf c}, Naphthalen-based Hopf-linked molecule. A building block to construct a chain using a benzene ring.
In this case, we expect $5 \times 5=25$ configurations for each Hopf-linked molecule, depending on the relative angle and the position of Naphthalene molecules. 
	{\bf d}, Cyclodecapentaene-based Hopf-linked molecule. The strains of bonds are significantly relaxed.
	{\bf e}, Accommodations of 2 and 4 rings into cyclodecapentaene-based Hopf-linked molecule. 
	{\bf f}, Tetracene-based Hopf-linked molecule. Graphene nano-ribbons can also be used to construct a Hopf-links
}
\end{center}
\end{figure}

\pagebreak
\begin{figure}
\begin{center}
\includegraphics[width=13.6cm]{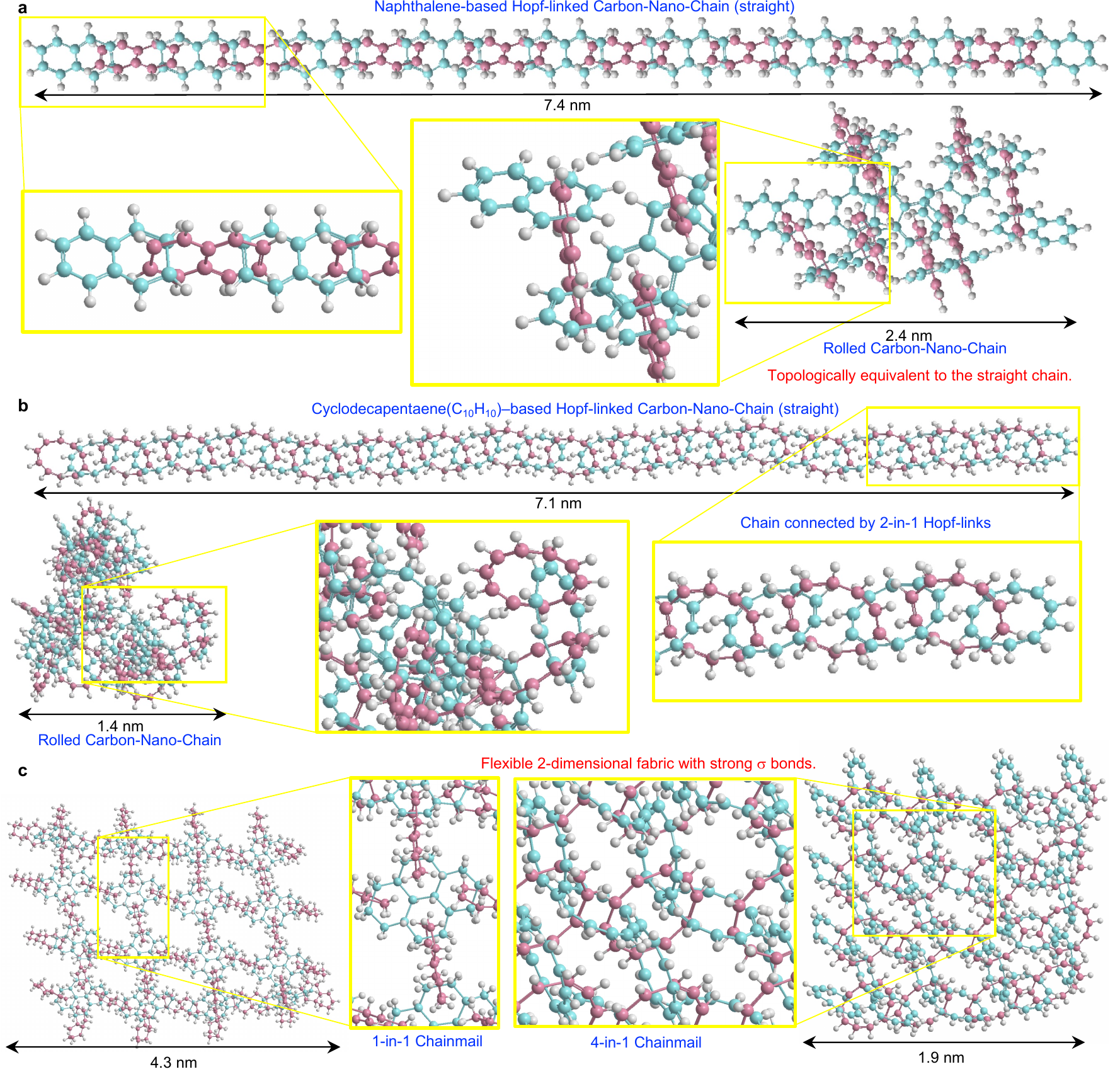}
\caption{
	{\bf  Carbon-Nano-Chains and Chainmail.} 
	{\bf a}, Naphthalene-based Hopf-linked chain. 
The chain can change the global geometries without changing the topology of the chain protected by strong $\sigma$ bonds. 
	{\bf b}, Cyclodecapentaene-based Hopf-linked chain. 
Straight and rolled chains are topologically the same, keeping the robust Topological-Long-Range-Order, while the translational symmetries are broken due to its flexible nature of the chain.
	{\bf c}, Chainmail using Hopf-links. Only 1 benzene ring can enter into another benzene ring, while cyclodecapentaene (C$_{10}$H$_{10}$) can accommodate the maximum of 4 rings. 
}
\end{center}
\end{figure}

\begin{figure}
\begin{center}
\includegraphics[width=12cm]{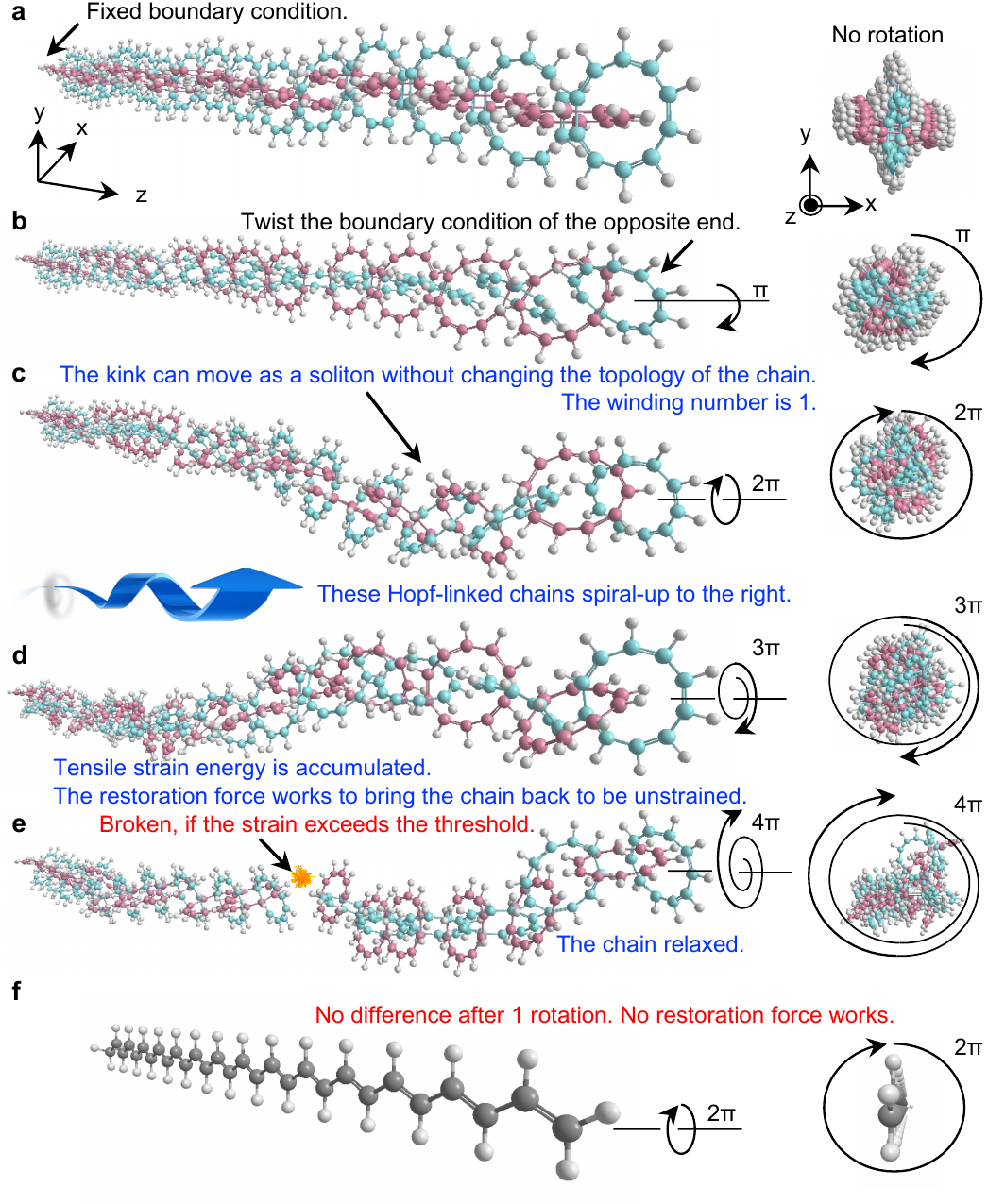}
\caption{
	{\bf  Stability of Hopf-linked Chain.} 
The stability of the chain was examined by imposing a fixed boundary condition.
	{\bf a}, Original straight-chain. No global strain was accumulated.
	{\bf b}, $\pi$-twisted chain. One of the ring at the boundary was $\pi$ rotated to the right.
	{\bf c}, $2\pi$-twisted chain. The winding number of the chain is 1, so that the kink appeared as a soliton, which is topologically protected upon deformations of the chain.
	{\bf d}, $3\pi$-twisted chain. The restoration forces worked to bring the chain back to the straight line.
	{\bf e}, $4\pi$-twisted chain. The chain was broken down and the strain was relaxed.
	{\bf f}, Poly-acetylene chain. No restoration force worked upon the rotation of $2\pi$ due to its rotational symmetry.
}
\end{center}
\end{figure}

\pagebreak
\begin{figure}
\begin{center}
\includegraphics[width=13.6cm]{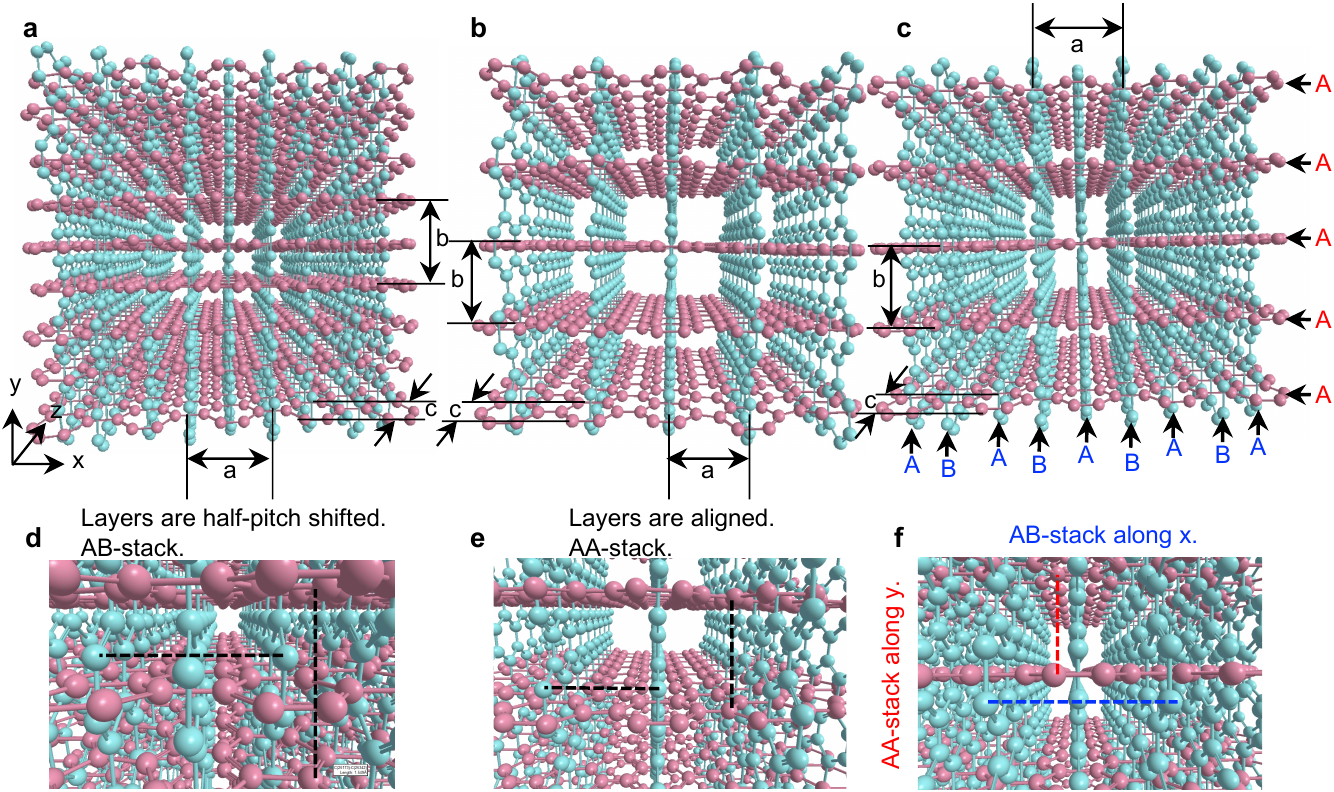}
\caption{
	{\bf  New $3D$ carbon allotrope, Hopfene.} 
	{\bf a}, (1,1) Hopfene with the maximum insertions of Graphene sheets in the unit cell with AB-stacks. 
It is a tetrahedral structure with the lattice constant $a=b$.
	{\bf b}, (2,2) Hopfene with aligned AA-stacks adjacent Graphene layers. It is also a tetrahedral structure with the lattice constant $a=b$.
	{\bf c}, (1,2) Hopfene with AB stacks along horizontal ($x$) direction and AA-stacks along vertical direction ($y$) . It is also a tetrahedral structure with the lattice constant $a=b$.
	{\bf d-f}, Expanded views of (1,1), (2,2), and (1,2) Hopfene crystals.
}
\end{center}
\end{figure}

\begin{figure}
\begin{center}
\includegraphics[width=15cm]{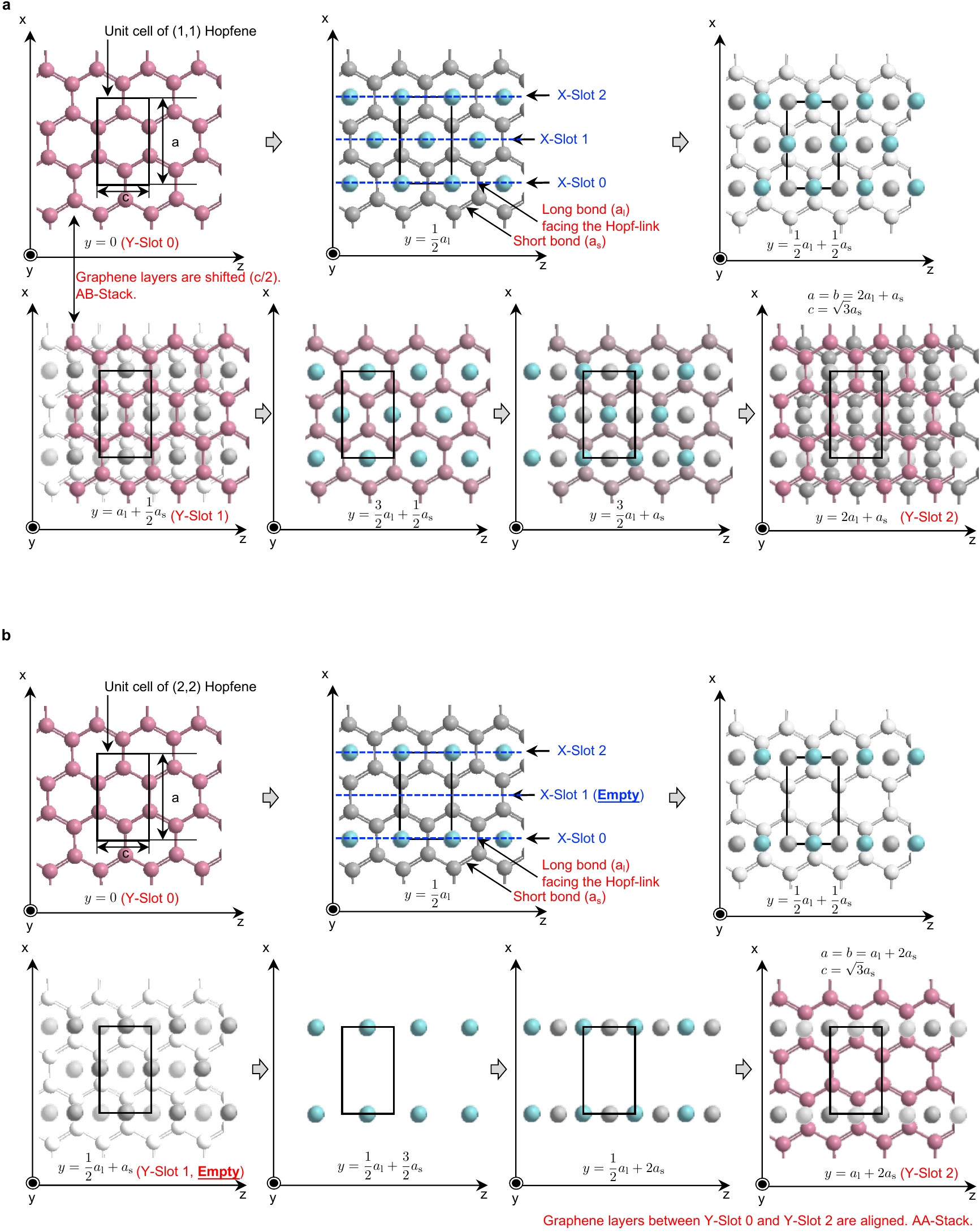}
\end{center}
\end{figure}
\pagebreak
\setcounter{figure}{5}
\begin{figure}
\begin{center}
\caption{
	{\bf  Unit cells for (1,1) and (2,2) Hopfene structures.} 
The crystal structure is tetragonal for both structures ($a=b\neq c$).
We show the details of atomic arrangements in unit cells upon increasing the layers.
The bond length of the bond across the Hopf-link was expanded to be long $a_{\rm l}$, while the expansion of other bonds was shorter $a_{\rm s}$.
		{\bf a}, Unit cell for (1,1) Hopfene. 
All available slots are occupied by Hopf-links. 
We obtained $a_{\rm l}=1.8$ $\rm \AA$ and $a_{\rm l}=1.5$ $\rm \AA$.
The lattice constants were $a=b=2a_{\rm l}+a_{\rm s}=5.1$ $\rm \AA$ and $c=\sqrt{3}a_{\rm s}=2.6$ $\rm \AA$.
		{\bf b}, Unit cell for (2,2) Hopfene. 
The middle slot (slot 1) in the unit cell is empty for both $x$ and $y$ directions.
We obtained $a_{\rm l}=1.8$ $\rm \AA$ and $a_{\rm l}=1.4$ $\rm \AA$.
The lattice constants are $a=b=a_{\rm l}+2a_{\rm s}=4.6$ $\rm \AA$ and $c=\sqrt{3}a_{\rm s}=2.4$ $\rm \AA$.
}
\end{center}
\end{figure}

\pagebreak
\begin{figure}
\begin{center}
\includegraphics[width=8.9cm]{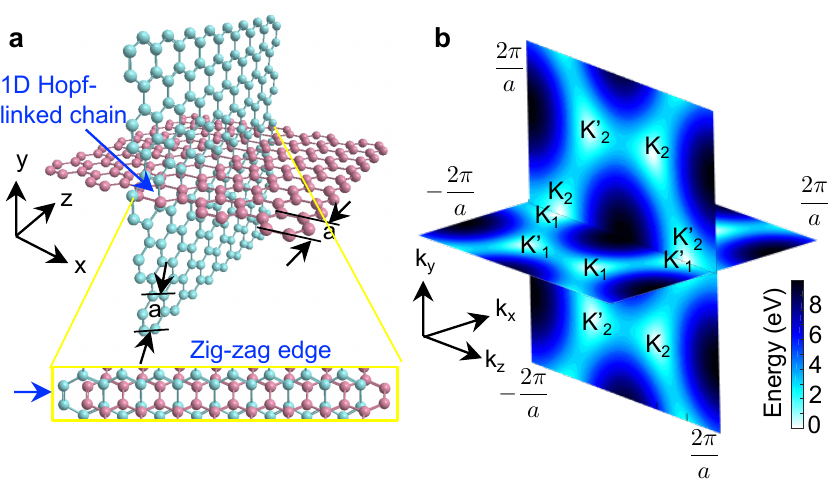}
\caption{
	{\bf  Hopf-linked bilayer-Graphene.} 
	{\bf a}, Simulated structure. $1D$ Hopf-linked chain was made at the intersection. The bonds were expanded in the chain, while other bonds away from the link was fully relaxed due to its open boundary condition during the simulation.
	{\bf b}, Expected electronic band structure, calculated by the tight-binding approximations. 
The horizontal Graphene has 2 valleys at K$_1$ and K$^{'}_1$, while the vertical one has 2 valleys at K$_2$ and K$^{'}_2$.
K$_1$ (K$^{'}_1$) and K$_2$ (K$^{'}_2$) are located at the equivalent position in the momentum space, so that states are completely degenerate at these points, while the propagating sheets in real space are different.}
\end{center}
\end{figure}

\pagebreak
\renewcommand{\figurename}{Supplementary Figure}
\setcounter{figure}{0}

\pagebreak
\begin{figure}
\begin{center}
\includegraphics[width=18.3cm]{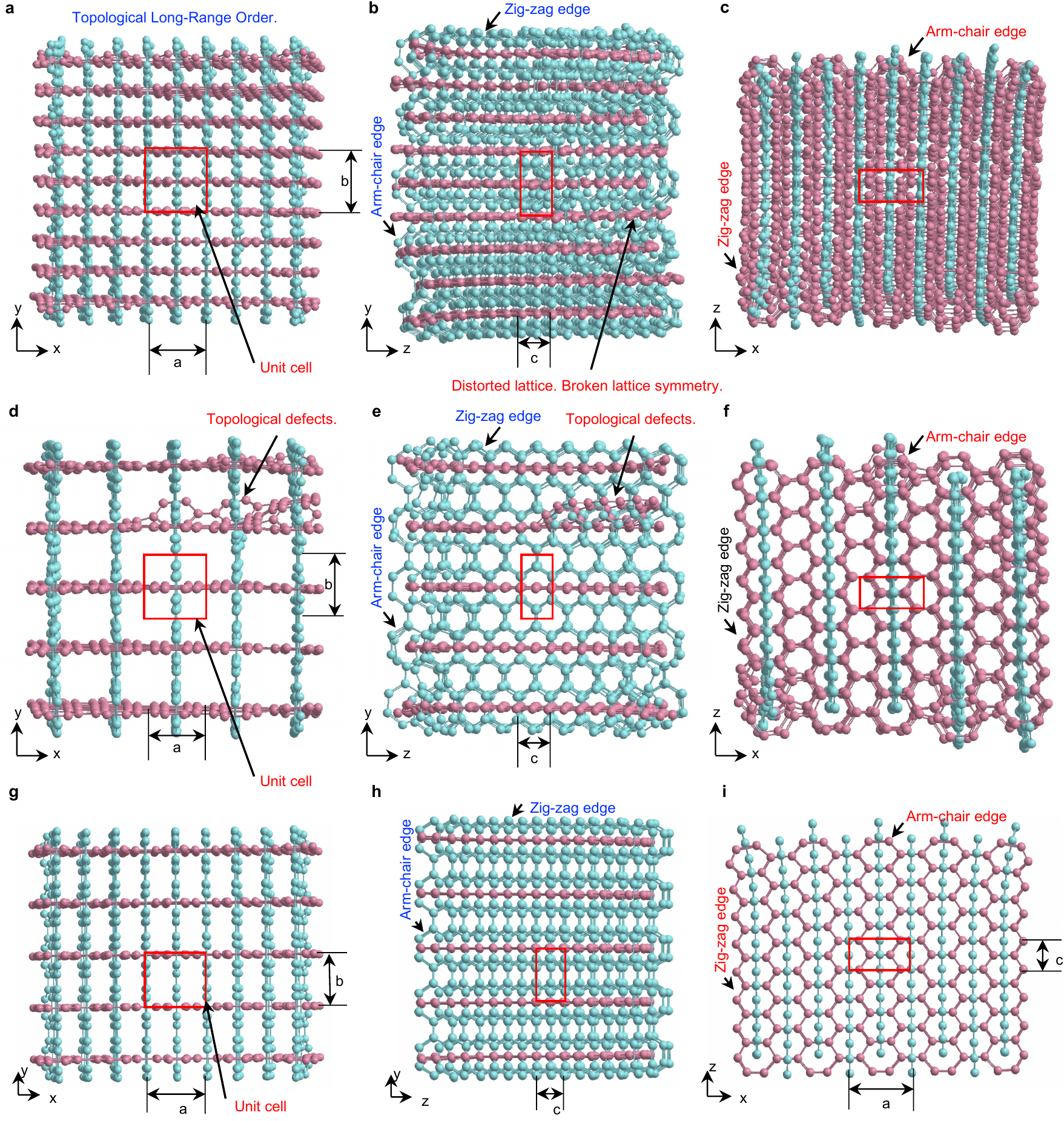}
\end{center}
\end{figure}
\pagebreak
\setcounter{figure}{0}
\begin{figure}
\begin{center}
\caption{
	{\bf Extended views of Hopfene.} 
Top views ({\bf a}, {\bf d}, {\bf g}), side views ({\bf b}, {\bf e}, {\bf h}), and the other side views ({\bf c}, {\bf f}, {\bf i}) of (1,1), (2,2), (1,2) Hopfene structures, respectively.
The Graphene sheets are distorted especially at the boundaries due to the relaxation of strains. 
Topological periodic orders were mostly maintained, while we also found topological defects in (2,2) Hopfene with broken topological periodicity.
This was coming from the non-perfect initial condition in our simulation, which could be corrected but we intentionally kept for highlighting the new type of defects without breaking bonds.
(1,2) Hopfene was perfect to keep the topological order.
}
\end{center}
\end{figure}

\pagebreak

\begin{figure}
\begin{center}
\includegraphics[width=14cm]{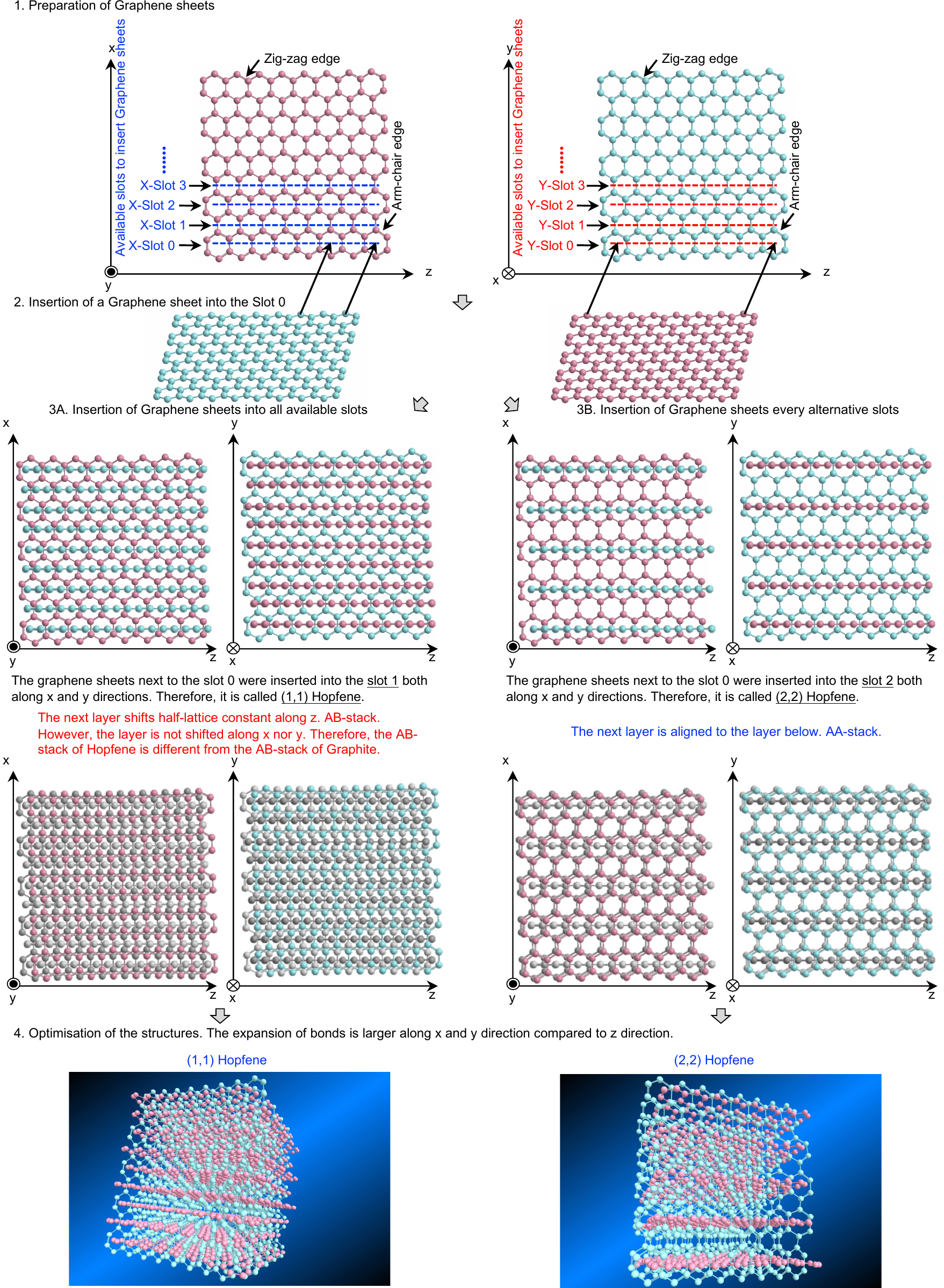}
\end{center}
\end{figure}
\pagebreak
\setcounter{figure}{1}
\begin{figure}
\begin{center}
\caption{
	{\bf  Classification of Hopfene crystal structures.} 
In the ideal crystal, we will insert infinite number of Graphene sheets into another stacks of infinite sheets. 
In the real experiments and simulations, the number of inserted sheets is finite, and we can classify the crystal by considering how to insert Graphene sheets.
We can assign available {\it slots} in Graphene sheets both horizontally ($x$) and vertically ($y$), and assume that we will insert the sheets in the slot 0.
If we insert next sheet to the nearest available slot, which is the slot 1 both for$x$ and $y$ directions and repeat the insertion process, and then, we can construct (1,1) Hopfene structure.
If we insert it into the slot 2, similarly, we can construct (2,2) Hopfene structure.
}
\end{center}
\end{figure}

\pagebreak

\begin{figure}
\begin{center}
\includegraphics[width=18.3cm]{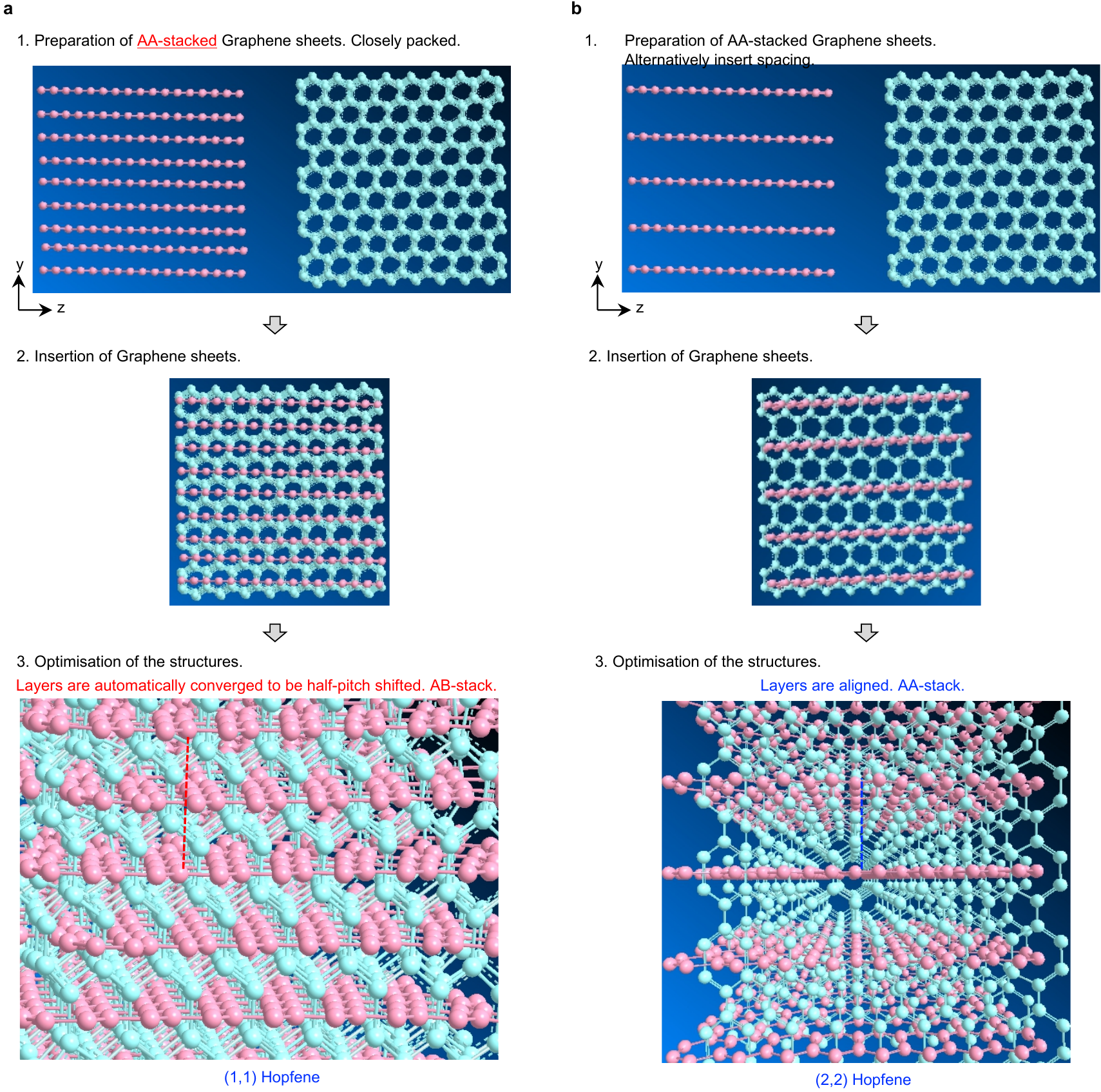}
\end{center}
\end{figure}
\begin{figure}
\pagebreak
\setcounter{figure}{2}
\begin{center}
\caption{
	{\bf  Procedure of simulations to make Hopfene crystal structures.} 
		{\bf a}, Steps for (1,1) Hopfene.
		{\bf b}, Steps for (2,2) Hopfene.
We have used AA-stacked Graphene sheets for simulations of both structures. 
First, horizontally AA-stacked Graphene sheets and vertically AA-stacked Graphene sheets were prepared. 
Fully relaxed Graphene structure was used for each sheet and the sheet was placed to the same distance with the lattice constant of Graphene.
If we see the original structure from the other side, vertical stacks of Graphene sheets (blue) were identified (not shown).
Then, we inserted the horizontal Graphene sheets (red) into vertical ones (blue, step2), and optimised the structure.
It was automatically converged to the appropriate stacks by minimising the energy.
}
\end{center}
\end{figure}

\pagebreak

\begin{figure}
\begin{center}
\includegraphics[width=7cm]{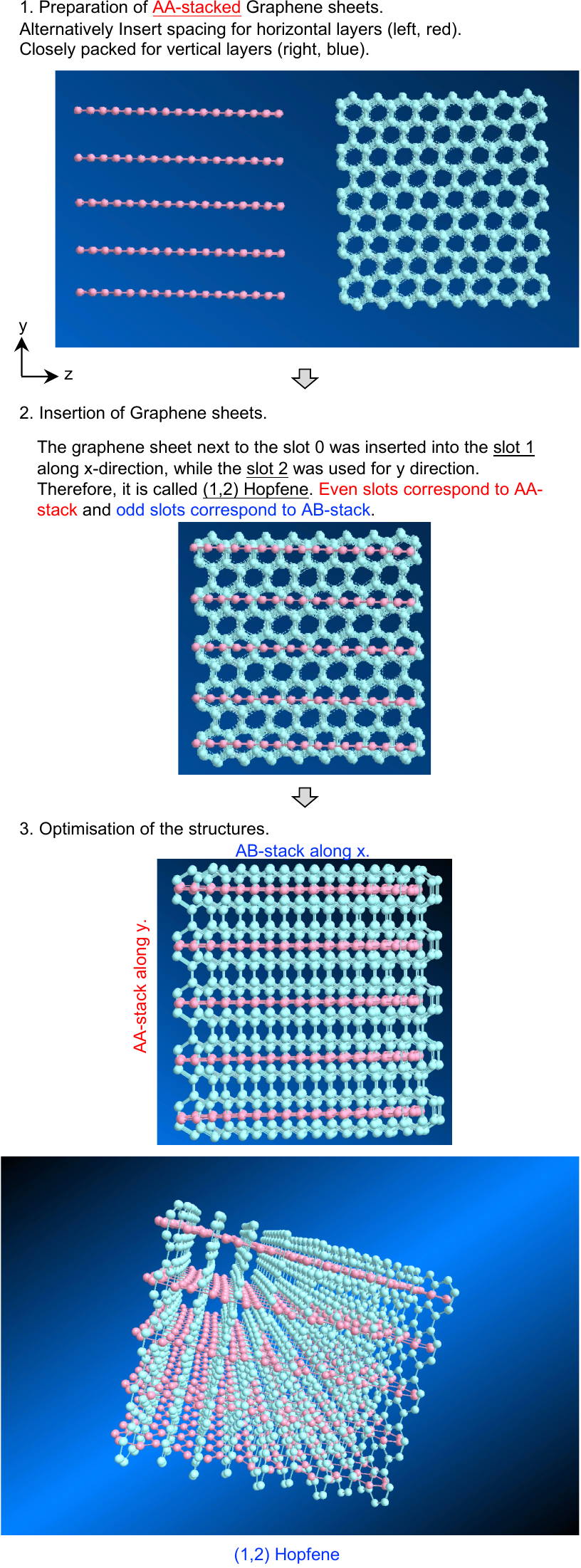}
\end{center}
\end{figure}
\pagebreak
\setcounter{figure}{3}
\begin{figure}
\begin{center}
\caption{
	{\bf  Procedure of simulation to make (1,2) Hopfene crystal structure.} 
We have also used AA-stacked Graphene sheets for both vertical and horizontal stacks. 
Similar to (1,1) and (2,2) Hopfene simulations, we have prepared appropriate number of sheets, while keeping appropriate distance.
The slot 1 is used for $x$-direction, while it is empty for $y$-direction (red).
By optimising the structure, we found that AB-stacking along $x$-direction, while it was AA-stacking along $y$-direction, as we expected.
}
\end{center}
\end{figure}

\pagebreak

\begin{figure}
\begin{center}
\includegraphics[width=18.3cm]{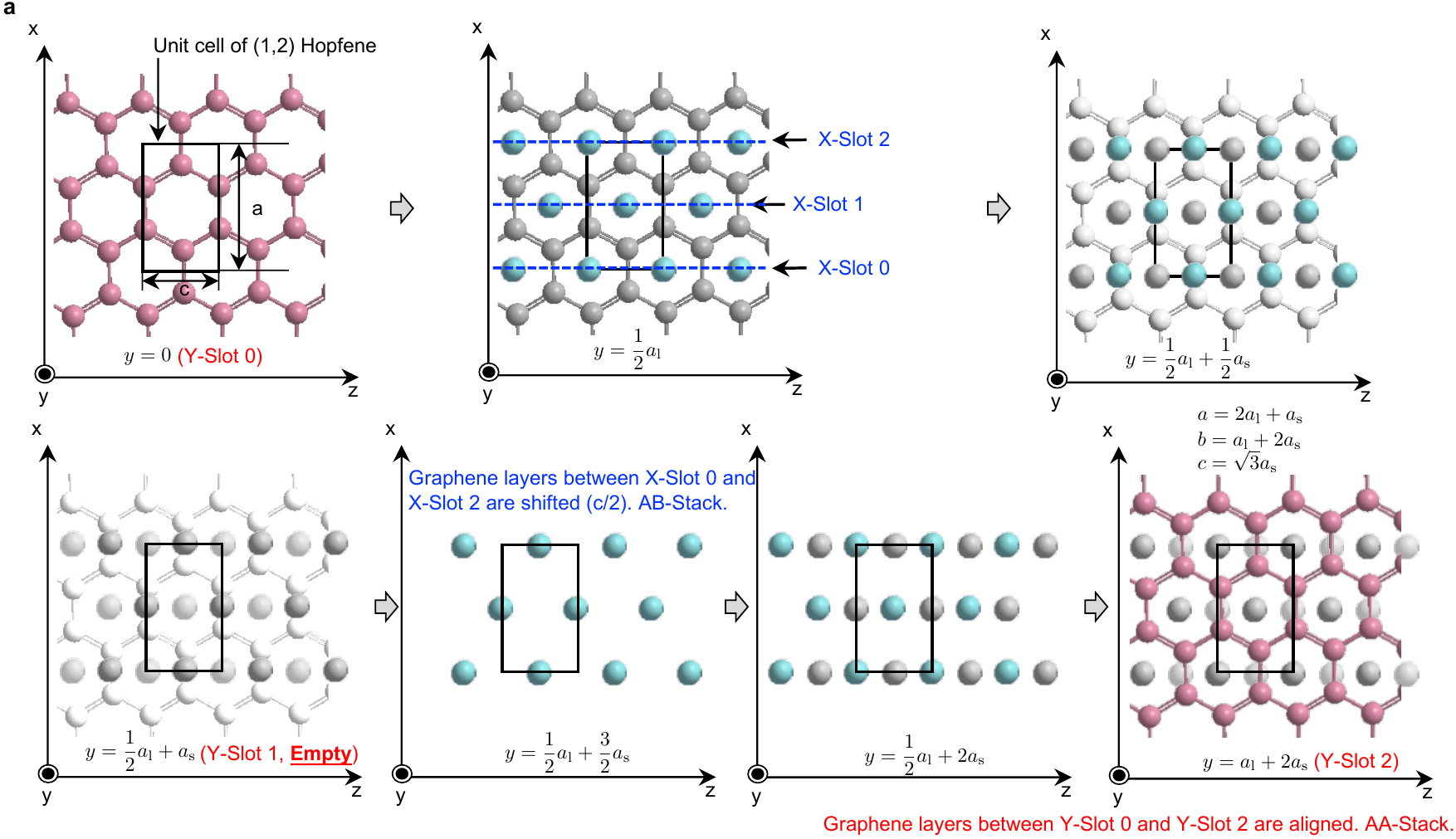}
\caption{
	{\bf  Unit cell for (1,2) Hopfene.} 
All available slots are occupied by Hopf-links along $x$-direction, while the middle slot (slot 1) is empty along $y$ direction.
Therefore, the crystal structure is orthorhombic ($a\neq b \neq c$). 
We obtained $a_{\rm l}=1.8$ $\rm \AA$ and $a_{\rm l}=1.5$ $\rm \AA$.
The lattice constants were $a=2a_{\rm l}+a_{\rm s}=5.1$ $\rm \AA$, $b=a_{\rm l}+2a_{\rm s}=4.8$  $\rm \AA$, and $c=\sqrt{3}a_{\rm s}=2.6$  $\rm \AA$.
}
\end{center}
\end{figure}

\pagebreak

\begin{figure}
\begin{center}
\includegraphics[width=15cm]{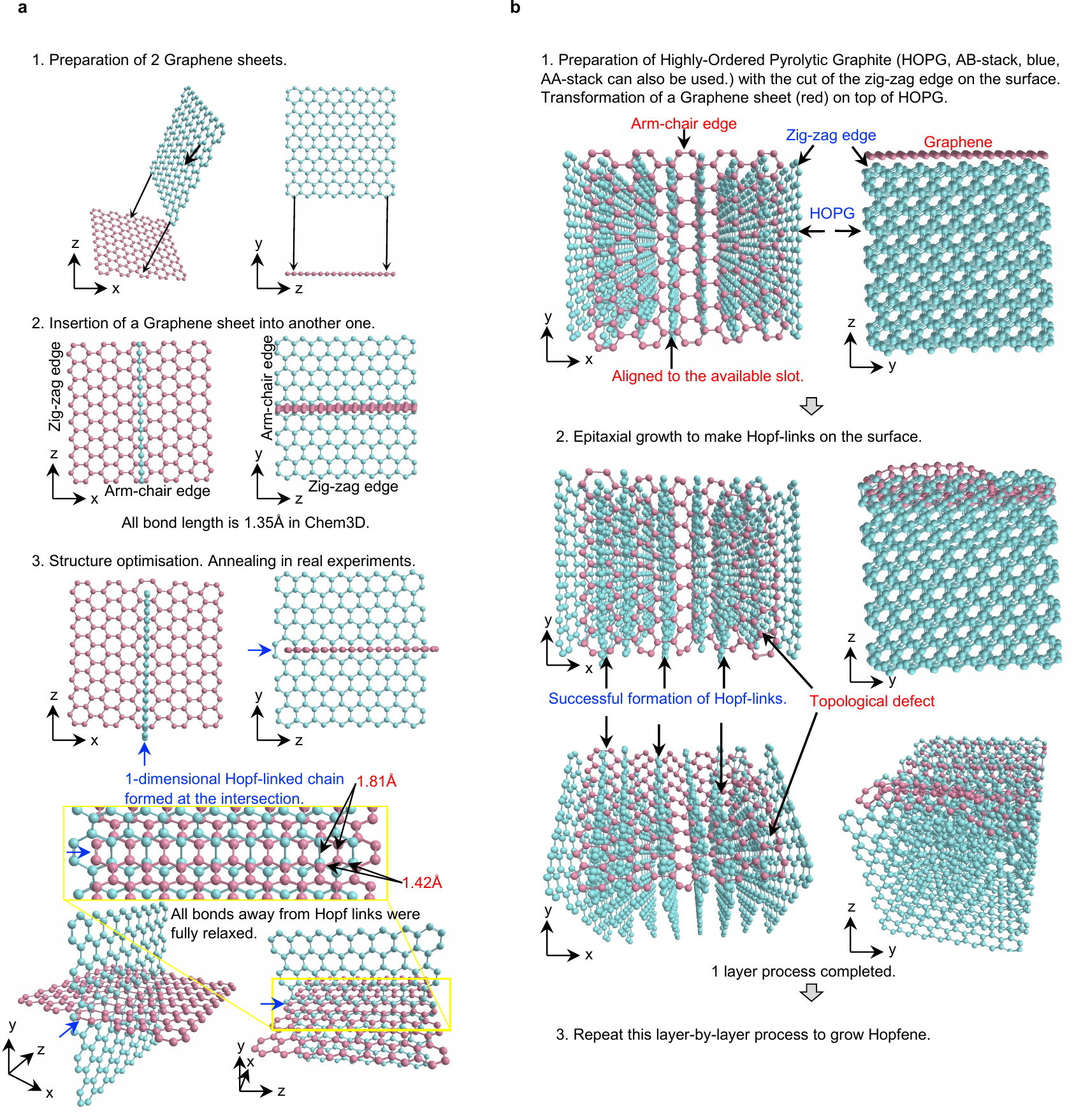}
\end{center}
\end{figure}
\pagebreak
\setcounter{figure}{5}
\begin{figure}
\begin{center}
\caption{
	{\bf  Proposed procedure to make Hopf-linked bilayer-Graphene.} 
{\bf a}, Simulated structure upon the vertical insertion of a Graphene sheet to the other one. 
$1D$ Hopf-linked Carbon-Nano-Chain was formed at the intersection.
Bonds across the Hopf-links were highly strained, while the other bonds are relaxed.
{\bf b}, Possible experimental procedure to make Hopf-links. 
The Graphene sheet will be exfoliated on top of the Highly-Oriented-Pyrolytic-Graphite (HOPG) with the zig-zag edge on the front surface.
The pitch of the zig-zag edge is the same with the pitch of the empty slots, if the Graphene sheets are aligned perpendicular to the HOPG layers.
After the transfer, the epitaxial growth technique will be employed to grow Graphene bonds to close the Hopf-links.
The transferred Graphene sheets are highly strained after the formation of Hopf-links. 
By repeating this layer-by-layer transfer and growth techniques, the Hopfene crystal will be grown.
}
\end{center}
\end{figure}

\pagebreak

\begin{figure}
\begin{center}
\includegraphics[width=18.3cm]{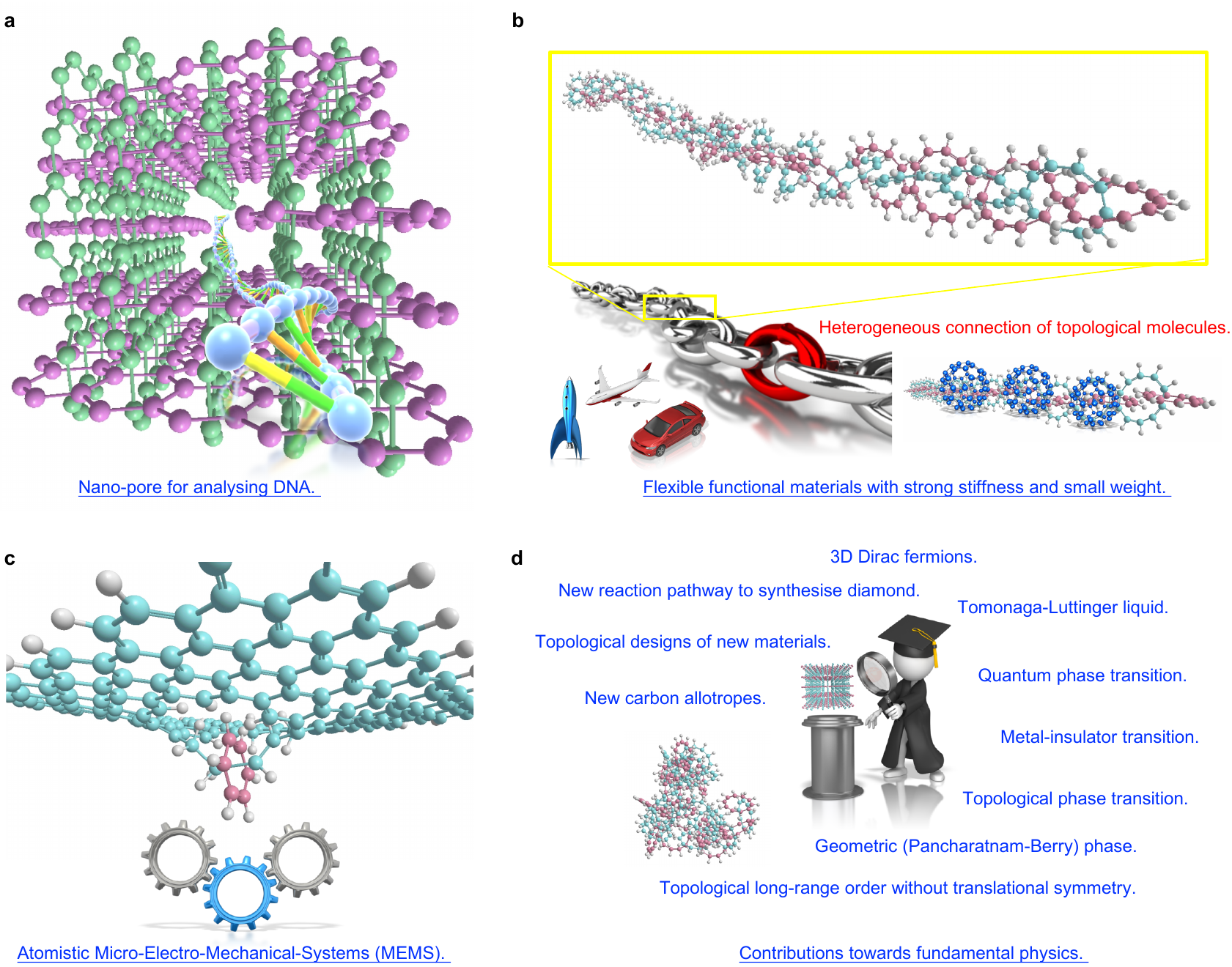}
\caption{
	{\bf  Possible applications of topological materials.} 
{\bf a}, Nano-pore for DNA detection. The size of the pore will be controlled in the atomic scale, while the crystal will be mechanically strong.
{\bf b}, Flexible, strong, and light-weighted functional materials. We can also think about the heterogeneous links to other materials.
{\bf c}, Molecular motors for NEMS.
{\bf d}, Contributions towards material science and physics.
}
\end{center}
\end{figure}

\end{document}